\documentclass[aps,pre,preprint,twocolumn,english,a4paper,10pt,floatfix,showpacs]{revtex4-1}
\usepackage[T1]{fontenc}
\usepackage{url}
\usepackage[dvips]{graphicx}
\usepackage{amsmath,amsfonts,amssymb,amsbsy}  

\newcommand{\bb}{\mathbf}

\newcommand{\half}{\frac{1}{2}}

\begin{document}
\title{Thermal conduction and interface effects in nanoscale Fermi--Pasta--Ulam conductors}
\author{K. S\"a\"askilahti}
\email{kimmo.saaskilahti@aalto.fi}
\author{J. Oksanen}
\author{R.~P. Linna}
\author{J. Tulkki}
\affiliation{Department of Biomedical Engineering and Computational Science, Aalto University, FI-00076 AALTO, FINLAND}
\date{\today}

\begin{abstract}
 We perform classical non-equilibrium molecular dynamics simulations to calculate heat flow through a microscopic junction connecting two larger reservoirs. In contrast to earlier works, we also include the reservoirs in the simulated region to study the effect of the bulk-nanostructure interfaces and the bulk conductance. The scalar Fermi--Pasta--Ulam (FPU) model is used to describe the effects of anharmonic interactions in a simple manner. The temperature profile close to the junction in the low temperature limit is shown to exhibit strong directional features that fade out when temperature increases. Simulating both the FPU chain and the two bulk regions is also shown to eliminate the non-monotous temperature variations found for simpler geometries and models. We show that with sufficiently large reservoirs, the temperature profile in the chain does not depend on the details of thermalization used at the boundaries.  
\end{abstract}
 \maketitle

\section{Introduction}

Mesoscopic phonon transfer phenomena \cite{chen} have lately gained much attention due to the advances in fabrication and characterization of nanostructures and their potential in allowing the tailoring of thermal properties of materials \cite{cahill03}. Recent theoretical and experimental studies have allowed obtaining new insight and evidence of phononic phenomena such as ballistic transport \cite{berber00,kim01}, confinement effects \cite{rego98,schwab00}, interference \cite{prasher07b} and tunneling \cite{prunnila10,altfeder10}. The ability to manipulate heat flow  in microscopic level is expected to enable e.g. the design of new materials for thermoelectric conversion \cite{majumdar04,dubi11}, improved thermal management of future electronic devices \cite{pop10} or even information processing by phonons \cite{terraneo02, li06, chang06}.

In theoretical studies of thermal transport in nanostructures, two atomistic methods are especially popular. Landauer--B\"uttiker formalism \cite{landauer81,buttiker92} of phonon heat transfer \cite{angelescu98,rego98,blencowe99} can produce the frequency-dependent phonon transmission functions for nearly arbitrary geometries with little effort and is therefore especially well suited for contact resistance studies. However, the inclusion of anharmonic effects, i.e. phonon-phonon scattering, is very challenging. In classical molecular dynamics (MD) simulations, on the other hand, the inclusion of scattering is trivial, but since the computational effort increases very rapidly with the number of particles, the simulation of macroscopic structures is out of reach. Capturing the effects of a frequency-dependent coupling between the phonon modes at the interface between the nanostructure and the bulk region by correlated stochastic dynamics \cite{segal03,segal08,zhou10} has been limited to phenomenological level and cannot include e.g. the effect of anharmonic interactions in the bulk phonon spectrum.

Our work aims at bridging the gap between the Landauer--B\"uttiker formalism and MD in that we include both the reservoirs and anharmonic potential in the MD simulations, thereby making it possible to account in detail for the coupling of the phonons in the nanostructure and in the bulk and to access e.g. the local temperature profiles not only in the junction area, but also in the bulk. In contrast to previous molecular dynamics simulations of heat transfer through an orifice between two bulk silicon reservoirs \cite{saha07}, our model incorporates a wire-like nanostructure acting as a bridge between the two reservoirs. Our setup also differs from earlier simulations of carbon nanotubes on substrate \cite{diao08,hu08b,fan09,gao11} in that we do not supply a constant heat current in the hot and cold regions of the structure in the spirit of Jund and Jullien \cite{jund99}. Their method circumvents the problem of strongly fluctuating heat currents \cite{lukes00} seen when a constant temperature difference is 
imposed in the structure, making the calculation of thermal conductivity easier. However, the artificial supplement of heat in the middle of the junction obstructs phonon flow and one cannot obtain realistic temperature profiles for the whole structure. Thermal boundary conditions imposed in our setup are also closer to an actual experimental setup.

In order to study the effects of anharmonic interactions in a simple manner, we use the Fermi--Pasta--Ulam (FPU) potential \cite{fermi55} as the interatomic potential. The FPU model has become the prototype of a simple model that exhibits anomalous transport in low-dimensional systems \cite{bonetto00,lepri03,dhar08} but seems to be sufficiently refined to generate regular diffusive heat transfer in three dimensions \cite{saito10,wang10}. Our simulations allow one to understand in detail how e.g. temperature, geometry and the bulk phonons affect heat transfer in the nanostructure. The main drawback of MD is that quantum mechanical effects such as the reduced population of high-energy modes (compared to classical equipartition) cannot be trivially included.

The paper is organized as follows. In Sec. \ref{sec:model}, we present our computational setup and define various structural parameters. Section \ref{sec:results1} discusses temperature and heat current profiles in a setup consisting of two bulk regions and a contact constriction, focusing especially on the effect of phonon-phonon scattering. Section \ref{sec:results2} is devoted to studying how the non-equilibrium temperature profile and the temperature dependence of thermal conductance in an FPU chain connecting two FPU square lattices are altered compared to simpler thermalization. In Sec. \ref{sec:discussion}, we give some general remarks on the model and our results. Section \ref{sec:conclusions} concludes the work.

\section{Computational model}
\label{sec:model}
The molecular dynamics model is based on describing the dynamics and interactions of particles by a Hamiltonian
\begin{equation}
 H = \sum_i \frac{p_i^2}{2m} + \sum_{\langle i,j \rangle} V(z_i-z_j),
\label{eq:hamiltonian}
\end{equation}
where the first sum is over all the particles and the latter sum over pairs of particles that are located at neighboring sites. Momentum $p_i$ is conjugate to the scalar displacement co-ordinate $z_i$, $m$ is the mass of the atoms and $V$ is the interaction potential between particles. In the $\beta$-FPU model, the interatomic potential is taken to be (see, e.g., Ref. \cite{berman05})
\begin{equation}
 V(z)=\frac{1}{2}m\omega_0^2 z^2 + \frac{\beta}{4} z^4,
\label{eq:vdef}
\end{equation}
where $\beta$ is the anharmonicity parameter and $\omega_0$ is the harmonic oscillation frequency that sets the natural time scale for the system (see below). 

\begin{figure}
 \begin{center}
  \includegraphics[width=8.6cm]{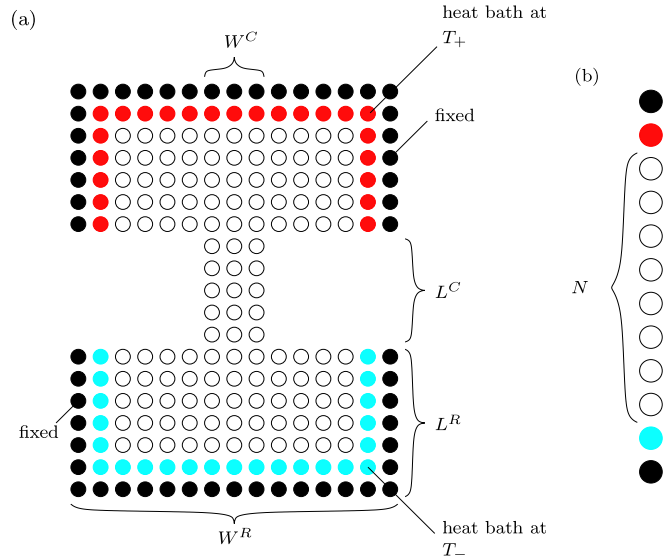}
 \caption{(Color online) (a) The square lattice model and the device setup. The positions of the black atoms are fixed, red (dark gray in black and white) atoms are in a Langevin bath at temperature $T_+$, cyan (light gray) atoms are in a bath at temperature $T_-$ and the white atoms are free (apart from interactions with neighbors). Atoms located at neighboring sites are connected by anharmonic springs. The size of the reservoirs is defined by the parameters $W^R$ and $L^R$ and the size of the constriction by $W^C$ and $L^C$. (b) A bath-terminated chain of $N$ atoms.}
 \label{fig:geom1}
 \end{center}
\end{figure}

We study heat transfer in a square lattice depicted in Fig. \ref{fig:geom1}(a), as well as in an atom chain of Fig. \ref{fig:geom1}(b). The black atoms in Fig. \ref{fig:geom1} are fixed in position and red and cyan atoms are set in heat baths at temperatures $T_+$ and $T_-$, respectively. It is most natural to imagine the coordinate $z_i$ to describe the out-of-plane displacement of a vibrating membrane \cite{fetter} at the lattice site $i$. It would be straightforward to consider more realistic lattices with more degrees of freedom, but this simple model is sufficient for studying qualitatively the heat transfer through an anharmonic nanostructure. Similar scalar FPU model has been used earlier in Refs.  \cite{saito10,wang10} in three spatial dimensions. 

To model isothermal boundary conditions, we use stochastic Langevin dynamics \cite{weiss} to model the motion of particles that are considered to be coupled to an external bath. The full equations of motion following from Eqs. \eqref{eq:hamiltonian}, \eqref{eq:vdef} and the additional Langevin noise are given by \cite{weiss}
\begin{equation}
 m\ddot{z}_i = - \sum_{j\in A_i} \left(m\omega_0^2 (z_i-z_j)+\beta(z_i-z_j)^3\right)  + \Theta_i\times\left(\eta_i- \gamma \dot{z}_i\right),
 \label{eq:eom}
\end{equation}
where $\Theta_i=1$ and $\Theta_i=0$ for particles coupled and not coupled to the heat bath, respectively, and $\gamma$ is the Langevin friction parameter. For convenience, we define $A_i$ to be the set of particles that are neighbors of the particle $i$. In a square lattice, each $A_i$ contains four elements (expect at the edges). The random variable $\eta$ has zero average $\langle \eta_i(t)\rangle=0$ and the fluctuations satisfy \cite{weiss}
\begin{equation}
 \langle \eta_i(t_1)\eta_j(t_2) \rangle = 2 \gamma T_i \delta (t_1-t_2)\delta_{ij},
\end{equation}
where we have set $k_B=1$. The magnitude of fluctuations is fixed by the requirement that the particle $i$ is connected to an external thermally equilibrated harmonic reservoir at temperature $T_i$. We assume the noise to be Markovian, i.e. the spectral density to be Ohmic.  

To study temperature profiles, we calculate the local kinetic energy
\begin{equation}
 T_i^{kin}=m\dot{z}_i^2.
\label{eq:tkindef}
\end{equation}
Even though we refer to the kinetic temperature later simply as the ''temperature'', reaching a steady state does not mean that the system is in local thermal equilibrium.

The heat current between two sites can be calculated by determining the time rate of change for the local Hamiltonian and identifying the heat current \cite{hardy63}. We define the local Hamiltonian by dividing the interaction energy evenly between two interacting sites, i.e.
\begin{equation}
 h_i = \frac{p_i^2}{2m}+\frac{1}{2} \sum_{j\in A_i} V(z_j-z_i).
\end{equation}
By using the equation of motion \eqref{eq:eom} for $p_i=\dot{z}_i/m$ and dropping the stochastic and frictional forces since they are non-zero only at the far edges of the bulk, one gets for the time derivative
\begin{equation}
 \dot{h}_i = -\sum_{j \in A_i} J_{ij},
\end{equation}
where 
\begin{equation}
J_{ij}=\half V'(z_i-z_j)(\dot{z}_i+\dot{z}_j)
\end{equation} 
is the heat current flowing from site $i$ to site $j$. We define the local heat current to be divided evenly between the sites of atoms between which the heat flows, such that
\begin{equation}
 \bb{J}_i = \half \sum_{j\in A_i} J_{ij} \frac{\bb{x}_j-\bb{x_i}}{|\bb{x}_j-\bb{x_i}|},
\label{eq:heatcurrdef}
\end{equation}
where the unit vector specifies the direction of each heat current. 

Once one has determined the heat current $J(T,\Delta T)$ flowing through a cross section of the constriction, thermal conductance can be calculated by
\begin{equation}
 K=\lim_{\Delta T\rightarrow 0} \frac{J(T,\Delta T)}{\Delta T}.
\label{eq:KT}
\end{equation}
Here the mean temperature is $T=(T_++T_-)/2$ and the temperature difference $\Delta T=T_+-T_-$. In practice, we use a small temperature difference $\Delta T$ (such that transport remains in the linear response regime) and divide the heat current by $\Delta T$ to find $K(T)$.

The equations of motion can be scaled to the dimensionless form by defining new time and position variables such that $t'= \omega_0 t$ and $z'=z/a_0$, where $a_0=\sqrt{m\omega_0^2/\beta}$ is the distance at which the harmonic and anharmonic forces are equal. The scaled equation of motion becomes
\begin{equation}
 \ddot{z}_i' = -\sum_{j\in A_i} (z_i'-z_j')(1+(z_i'-z_j')^2) +\Theta_i \times \left( \eta_i'-\gamma' \dot{z}_i' \right),
\end{equation}
where the dots are now time derivatives with respect to $t'$, $\gamma'=\gamma/(m\omega_0)$ is the dimensionless friction constant and $\eta'$ satisfies
\begin{equation}
 \langle \eta_i'(t_2')\eta_j'(t_1') \rangle = 2\gamma' T_i' \delta(t_1'-t_2')\delta_{ij},
\end{equation}
where the dimensionless temperature is defined by $T_i'=\beta T_i/(m^2\omega_0^4)$. The scaling clearly reveals that the external bath temperatures and anharmonicity are not independent variables -- increasing the temperature is equivalent to increasing the anharmonicity. This is easy to understand intuitively by noting that when the bath temperatures are increased, the atoms vibrate further away from the equilibrium positions and the anharmonic term becomes more important. The only free variables left in the model are the friction parameter $\gamma'$ and the bath temperatures $T'$.

The equations of motion are integrated using the velocity Verlet algorithm \cite{allentildesley} with a time step $\Delta t=0.01$ that we have verified to be sufficiently small by comparing the results to simulations performed with the time step $\Delta t=0.005$ and confirming that the results agree up to statistical error. Langevin dynamics is modeled in a standard manner \cite{allentildesley}. The normally distributed random numbers required for the Langevin dynamics are generated using the ziggurat algorithm \cite{marsaglia00}. After an initial simulation time of $t_{eq}=O(10^6-10^7)$, a steady state has been reached and one can start collecting data. Each physical quantity $X$ is measured at intervals of $t=5$ for a time $t_{sim}=O(10^6-10^7)$, delivering the set of measurements $\{X_k\}_{k=1}^M$. The time average is then
\begin{equation}
 \langle X\rangle = \frac{1}{M} \sum_{k=1}^M X_k 
\end{equation}
and the standard deviation $\delta X$ of the mean is obtained by assuming uncorrelated readings \cite{allentildesley}:
\begin{equation}
 \delta X = \frac{\sigma_X}{\sqrt{M}},
\label{eq:varx}
\end{equation}
 where
\begin{equation}
 \sigma_X = \sqrt{\frac{1}{M} \sum_{k=1}^M \left(\langle X_k^2\rangle -\langle X \rangle ^2 \right)}
\end{equation}
is the standard deviation of $M$ readings. To describe the statistical uncertainty, we include error bars of length $5\times\delta X$ (98.8\% quantile) in the plotted data. Since all the quantities that we show in the results section are time averages, we drop the brackets in $\langle X\rangle$ and write simply $X$ everywhere.

In the following, we only use the dimensionless units and drop the primes. The dimensionless quantities can be scaled back to the original quantities using the replacements $t\rightarrow \omega_0 t$, $z\rightarrow z/a_0$, $T\rightarrow \beta T/(m^2 \omega_0^4)$, $J \rightarrow \beta J/(m^2\omega_0^5)$ and $K\rightarrow K/\omega_0$.

\section{Results}

\subsection{Temperature and heat current profiles}
\label{sec:results1}
\begin{figure}
 \begin{center}
  \includegraphics[width=8.6cm]{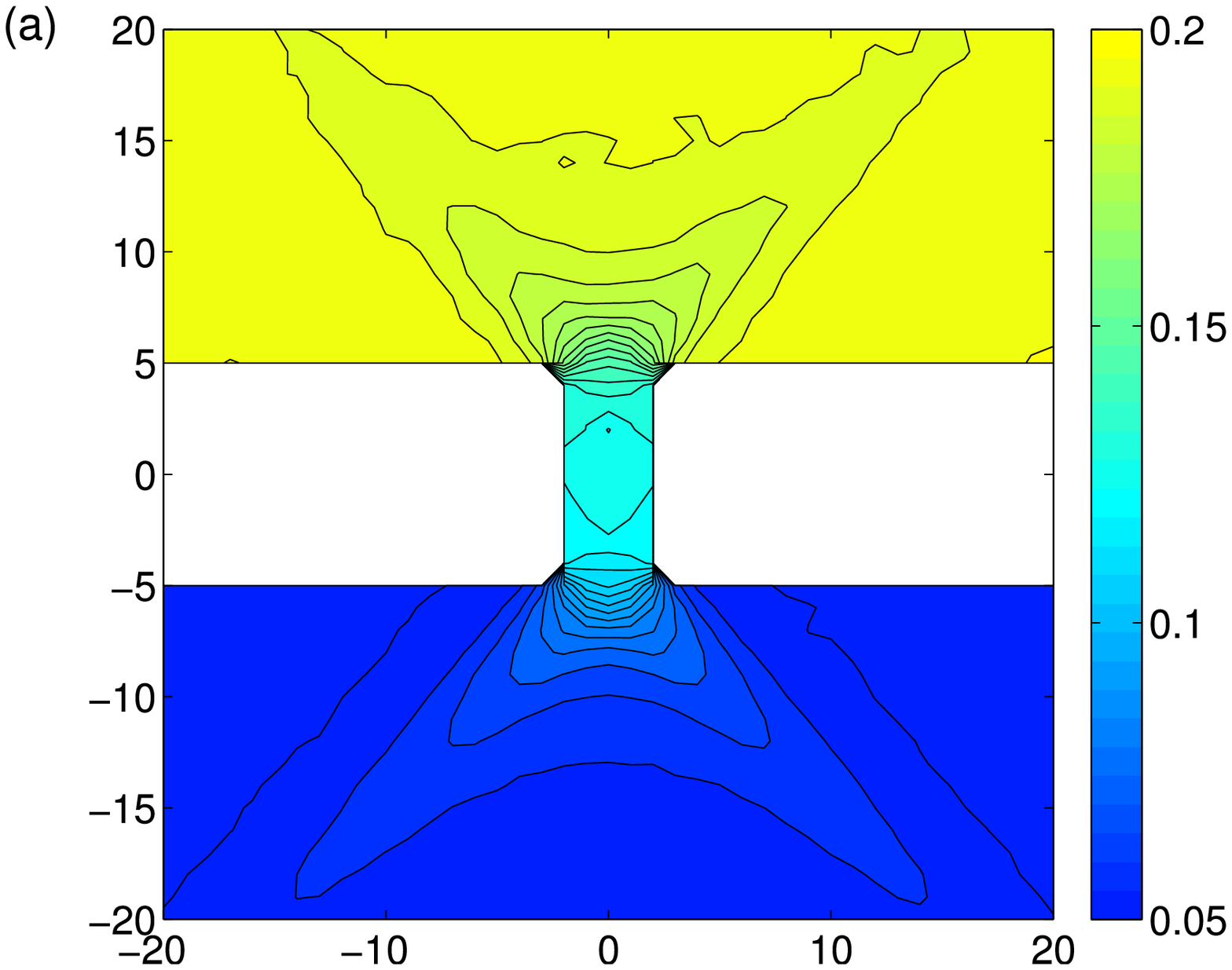}
  \includegraphics[width=8.6cm]{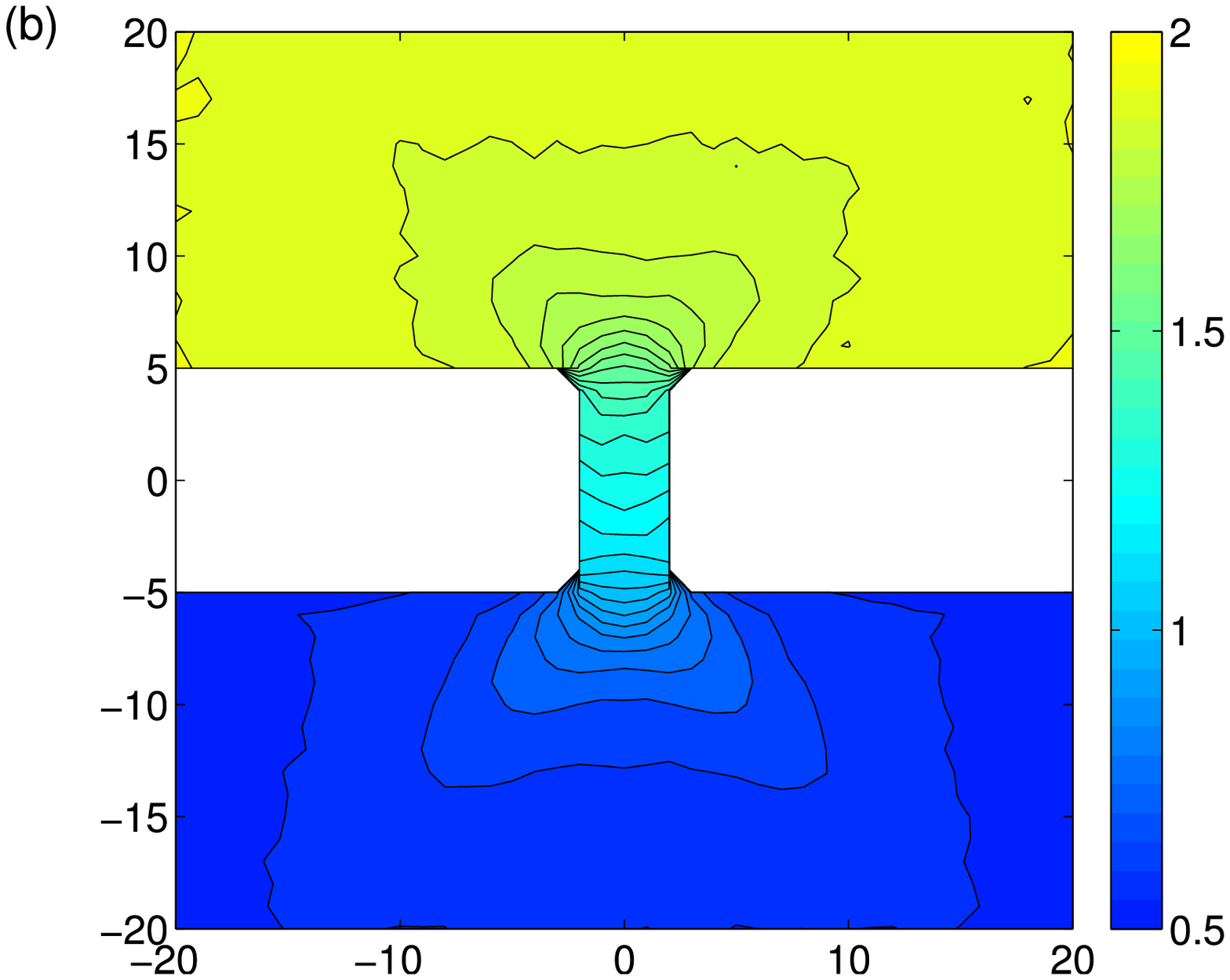}
 \caption{(Color online) Kinetic temperature profile at (a) low temperature ($T_+=0.20$, $T_-=0.05$) and (b) high temperature ($T_+=2.0$, $T_-=0.5$). The bulk size is $W^R=161$, $L^R=80$ and the constriction size $W^C=5$, $L^C=9$. The labels on the horizontal and vertical axes mark the atom indices. The separations of isolines are (a) $0.005$ and (b) $0.05$.}
 \label{fig:tprof1}
 \end{center}
\end{figure}

Figure \ref{fig:tprof1} shows the comparison of average kinetic temperature profiles (Eq. \eqref{eq:tkindef}) for a system with bulk size  $W^R=161$, $L^R=80$ and constriction size $W^C=5$, $L^C=9$ (see Fig. \ref{fig:geom1}) near the constriction region at (a) low temperature ($T_+=0.2$, $T_-=0.05$) and (b) high temperature ($T_+=2$, $T_-=0.5$). The Langevin friction parameter of the heat bath atoms is set to $\gamma=2$. We have obtained practically identical temperature profiles for larger systems and for periodic boundary conditions imposed in horizontal direction, suggesting that finite-size effects have been eliminated in the temperature profiles (not shown).

At low temperature (Fig. \ref{fig:tprof1}(a)), the temperature of the constriction region is nearly constant and temperature drops are concentrated at the interfaces between the bridge and the bulk due to contact resistance. The temperature profile in the bulk region exhibits directional dependencies that are also partly visible in the diamond-shaped feature in the constriction region. The directionality in the temperature profile at low temperature is expected to be a true physical effect related to the interference and mode-dependent coupling between the transversal modes in the bulk and in the constriction. It is important to stress that such wave features are strongly dependent on the symmetry of the lattice and therefore, more complicated lattices may exhibit very different directional dependencies in the temperature profiles. 

At higher temperature (Fig. \ref{fig:tprof1}(b)), phonon-phonon scattering is stronger and the low-temperature wave features described above fade out, leading to a more isotropic temperature profile in the bulk regions. Anharmonicity also increases the intrinsic resistance of the constriction, decreasing the relative contribution of the contact resistance to the total resistance and increasing the temperature gradient inside the constriction. Temperature drops nearly linearly inside the constriction. At even higher temperatures, the temperature profile becomes practically isotropic (not shown).

\begin{figure}
 \begin{center}
  \includegraphics[width=8.6cm]{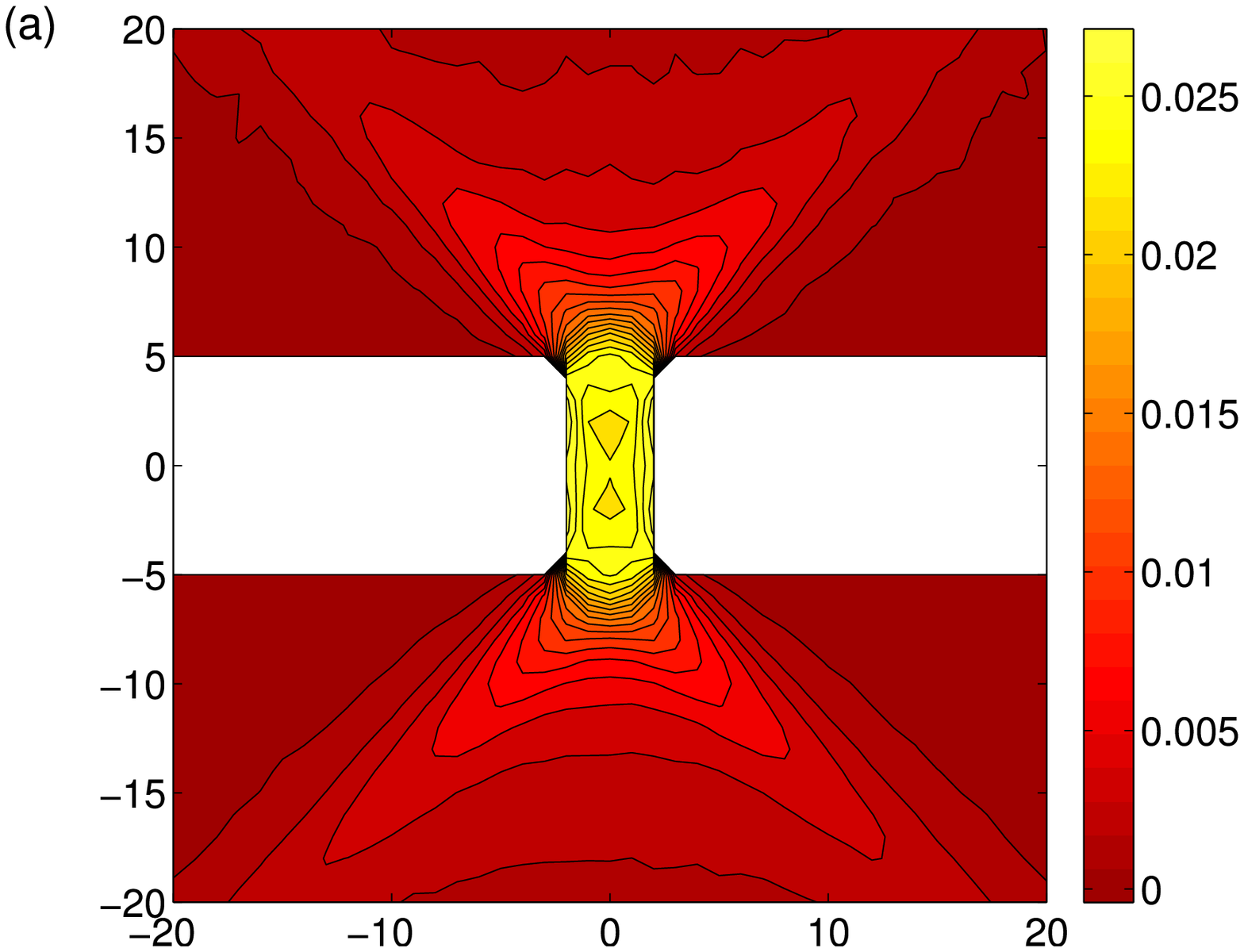}
  \includegraphics[width=8.6cm]{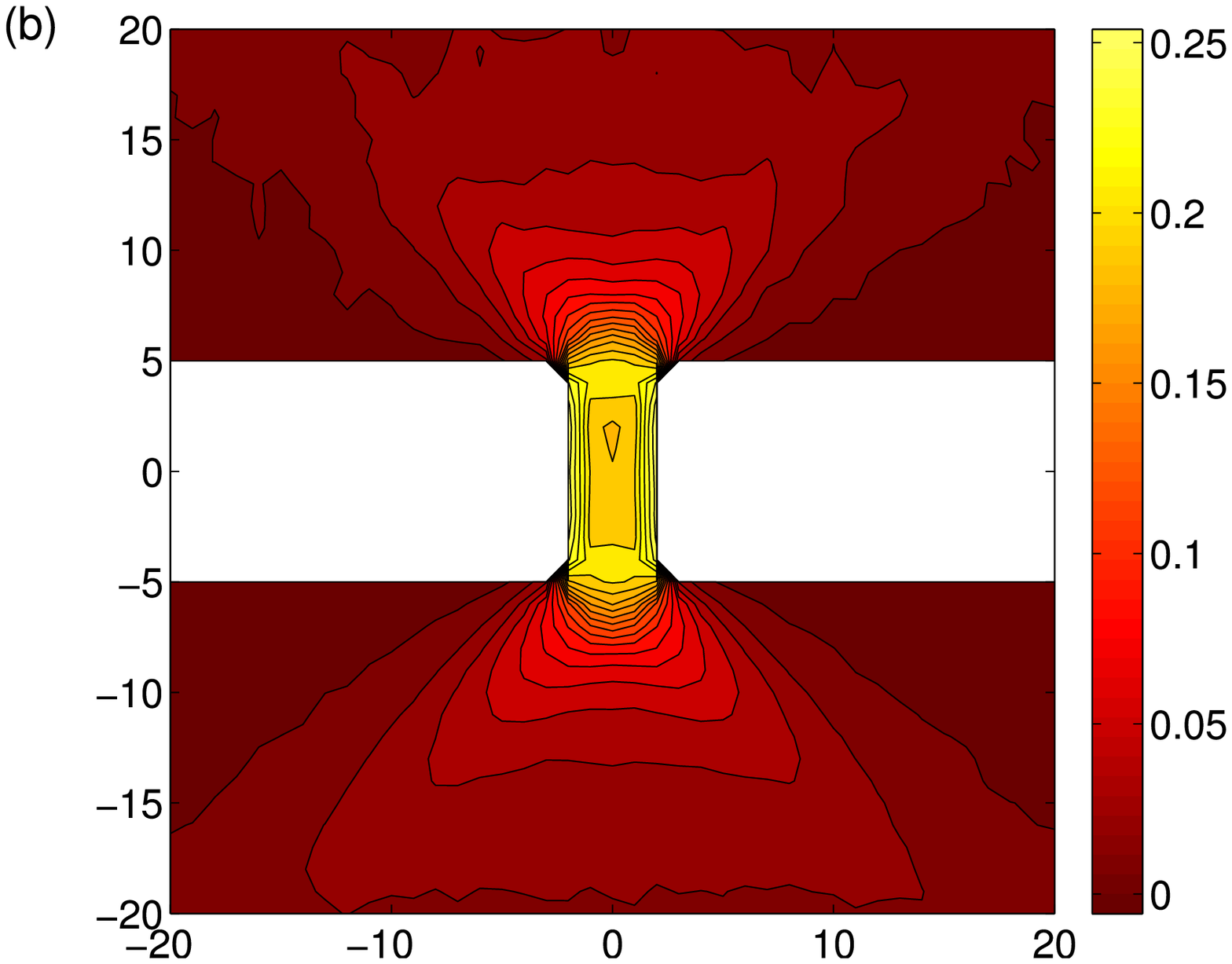}
 \caption{(Color online) Lateral local heat current $-J_i^y$ at (a) low temperature ($T_+=0.20$, $T_-=0.05$) and (b) high temperature ($T_+=2.0$, $T_-=0.5$). The labels on the horizontal and vertical axes mark the atom indices.}
 \label{fig:jprof1}
 \end{center}
\end{figure}

Figure \ref{fig:jprof1} shows the local heat currents (Eq. \eqref{eq:heatcurrdef}) corresponding to the temperature profiles of Fig. \ref{fig:tprof1}. We only show the negative vertical component $-J_i^y$ of the heat current. The transversal component also contains rich local variations, but since the component averages to zero at any horizontal cross-section, we concentrate on $J_i^y$. At both low (Fig. \ref{fig:jprof1}(a)) and high (Fig. \ref{fig:jprof1}(b)) temperature, heat flow is larger at the edges of the constriction than in the middle. The low temperature heat current profile exhibits strong interference features inside the constriction, reflecting again the fact that confined vibrational modes carry the heat. Furthermore, the heat current profiles show similar directional patterns as the temperature profiles in the bulk region. This effect is especially pronounced at low temperature, although some traces remain also at high temperature. Note that the numerical values in Figs. \ref{fig:jprof1}(a) and \ref{fig:jprof1}(b) differ by a factor of ten due to the ten-fold difference in the absolute temperatures.

Finding the sufficient bulk region size to eliminate bulk interference effects and discrete phonon spectrum of the thermal bulk reservoirs in the setup of Fig. \ref{fig:geom1} is one of the crucial computational parameters for the MD simulations of the extended system consisting of both bulk reservoirs and the constriction. The basic criteria for the sufficient size are that (i) the effect of the discrete mode spectrum of the finite bulk region becomes negligible and that (ii) changing the friction parameter does not change the temperature profile of the setup. Considering item (i), we examine in following the dependence of heat current on the bulk size. The dependence on friction parameter will be discussed in the next section.

\begin{figure}
 \begin{center}
  \includegraphics[width=8.6cm]{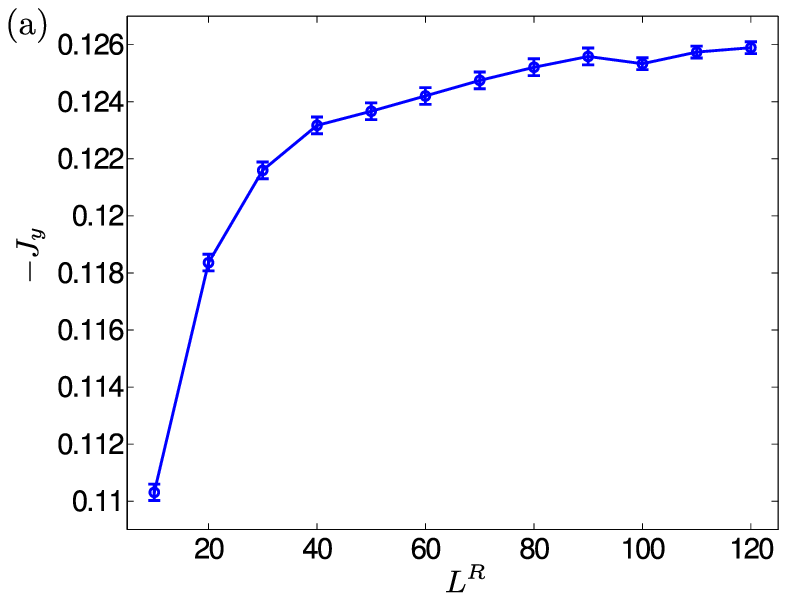}
  \includegraphics[width=8.6cm]{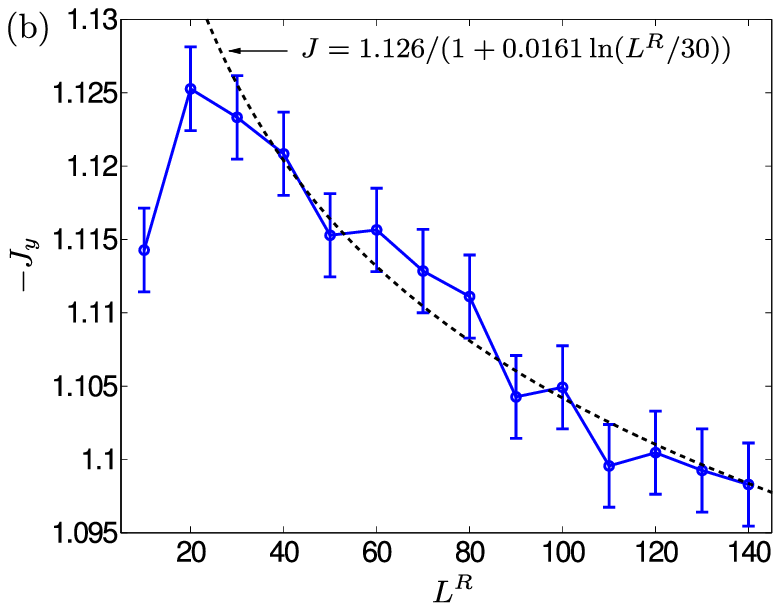}
 \caption{(Color online) Heat current flowing through a $W^C=5$, $L^C=9$ constriction as a function of the size of the reservoirs at (a) low temperature ($T_+=0.20$, $T_-=0.05$) and (b) high temperature ($T_+=2.0$, $T_-=0.5$). The transversal size of the bulk regions is $W^R=2L^R+1$. The heat currents flowing through the bridge have been averaged over the $L^C=9$ cross-sections of the bridge during the simulation and the error bars $5\delta$ were calculated in the end using Eq. \eqref{eq:varx}.}
 \label{fig:JNy}
 \end{center}
\end{figure}  

Figure \ref{fig:JNy} shows the total heat current flowing through the constriction as a function of the bulk length $L^R$ at (a) low temperature ($T_+=0.2$, $T_-=0.05$) and (b) high temperature ($T_+=2$, $T_-=0.5$). The width of the bulk is set to $W^R=2L^R+1$ and the size of the constriction is $W^C=5$, $L^C=9$ as before. At low temperature, the heat current increases monotonously as the bulk regions become larger and essentially saturates for large bulk sizes. This is easy to understand in the harmonic limit: increasing the transversal size of the bulk increases the number of transversal modes in the bulk. Since the phonons propagate without dissipation, this directly increases the conductance of the bulk and eventually the heat current is limited by the properties of the constriction only. These results suggest that at this low temperature range, the effect of the number of phonon modes on heat flow will be relatively small after the bulk size exceeds approximately $L^R\sim60$.

At high temperature, the dependence of the heat current on bulk size is more complicated. First of all, the error bars are larger, since the higher temperature induces more fluctuations. The shown error bars are very crude estimates for the fluctuations and should not be considered to be rigorous estimates. However, qualitative observations can be made. For $L^R=10$--$20$, the heat current increases as the bulk grows due to the increase in the number of modes as before. At $L^R\gtrsim 30$, the heat current starts to decrease due to increased resistance of the bulk region. This observation allows us to deduce that the relevant phonon scattering length in the system is of the order of tens of atomic distances in this temperature range. After $L^R\sim30$, the weak changes in heat current are expected to be mainly caused by the resistivity of the two-dimensional bulk, suggesting that for the high temperature range the bulk size needed to suppress finite-size effects is also of the order of few tens of atoms.

\begin{figure}
 \includegraphics[width=8.6cm]{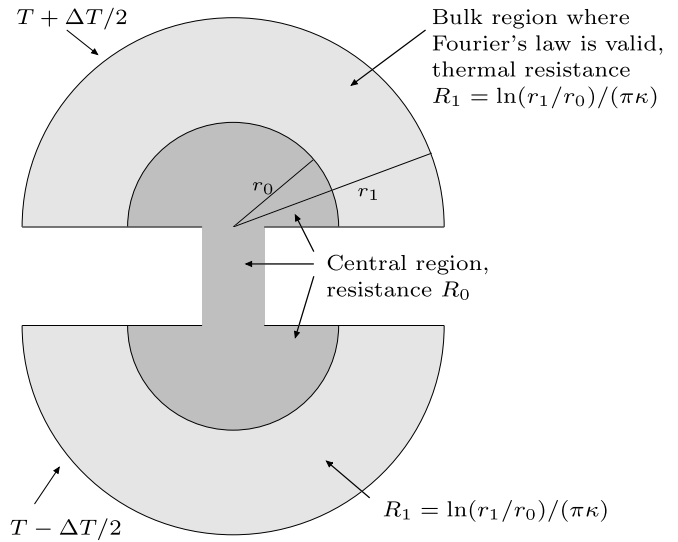}
 \caption{Schematic representation of the simple model used for explaining the dependence of heat current on the bulk size for large systems. The central region consisting of two discs of radius $r_0$ and the bridge contributes a constant factor $R_0$ the total thermal resistance. Heat transfer in the two bulk regions away from the central region is assumed to obey Fourier's law and therefore to contribute $R_1=\ln(r_1/r_0)/\pi\kappa$ to the thermal resistance. The current flowing through the structure is then given by Eq. \eqref{eq:bulk_model_0}.}
 \label{fig:4b_fit_geometry}
\end{figure}

To phenomenologically understand the dependence of the heat current on large bulk size in simple terms, we use a ''half-Fourier'' model depicted in Fig. \ref{fig:4b_fit_geometry}. The region to be studied is divided into three areas connected in series: The central region where Fourier's law does not need to hold and two bulk regions where heat transfer is approximately diffusive. The central region consisting of the bridge and two discs of radius $r_0$ is assumed to contribute a resistance $R_0$ to the total heat resistance $R=R_0+2\times R_1$. The bulk regions both contribute a resistance $R_1=\ln(r_1/r_0)/(\pi\kappa)$, where $\kappa$ is the thermal conductivity. This expression follows from solving the heat equation for two equiconcentric discs of radii $r_0$ and $r_1$. The total heat current as a function of the bulk radius $r_1$ is then
\begin{alignat}{2}
 J(r_1) &= \frac{\Delta T}{R_0+2\ln(r_1/r_0)/(\pi\kappa)}  \label{eq:bulk_model_0} \\
  &= \frac{J_0}{1+C\ln(r_1/r_0)}, 
 \label{eq:bulk_model}
\end{alignat}
where $J_0=\Delta T/R_0$ and $C=2/(\pi\kappa R_0)$ can be considered as fitting parameters. Identifying $L^R$ heuristically with $r_1$, we set $r_0=30$ and fit the heat current values obtained for $L_R=30,\dots,140$ to Eq. \eqref{eq:bulk_model}. Least-squares minimization delivers the fit $J_0\approx 1.126$ and $C\approx 0.0161$, and the corresponding line is shown dashed in Fig. \ref{fig:JNy}(b). While we acknowledge that the two-dimensional FPU model, like its one-dimensional counterpart, most likely exhibits anomalous thermal conductivity \cite{lepri03}, this simple model seems sufficient for explaining the dependence of the heat current on the bulk size for large bulk baths, suggesting that diffusive heat transfer is dominant in Fig. \ref{fig:JNy}(b). We note that in a three-dimensional system, in contrast to two-dimensional systems, Fourier's law predicts that the heat current should saturate to a constant for large bulk sizes. Since heat transfer in three-dimensional FPU model is diffusive \cite{saito10}, we expect that similar simulations for three-dimensional bulk regions would give a converging heat current as a function of the bulk size. Considering the long computational times required even for our two-dimensional system, we leave studies of the three-dimensional bulk system for future work. For the largest systems discussed in this paper ($\approx 80000$ particles), each simulation took approximately three days of wall time on 48 CPU cores in a local supercomputing cluster. It has recently been suggested \cite{roy12} that replacing the conventional FPU potential by a double-well potential would suppress the long-wavelength modes and allow reducing the system size and computation times in FPU-like systems.

\subsection{FPU chain}
\label{sec:results2}
\begin{figure}
 \begin{center}
 \includegraphics[width=8.6cm]{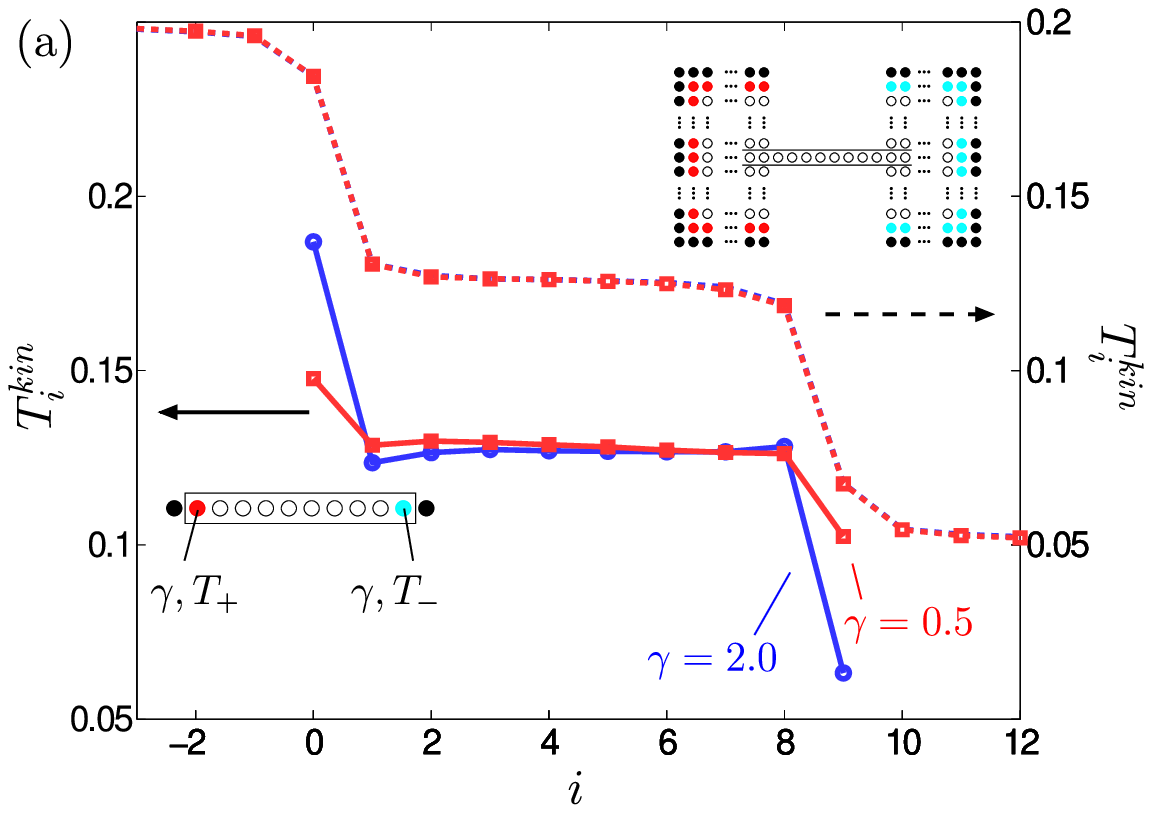}
 \includegraphics[width=8.6cm]{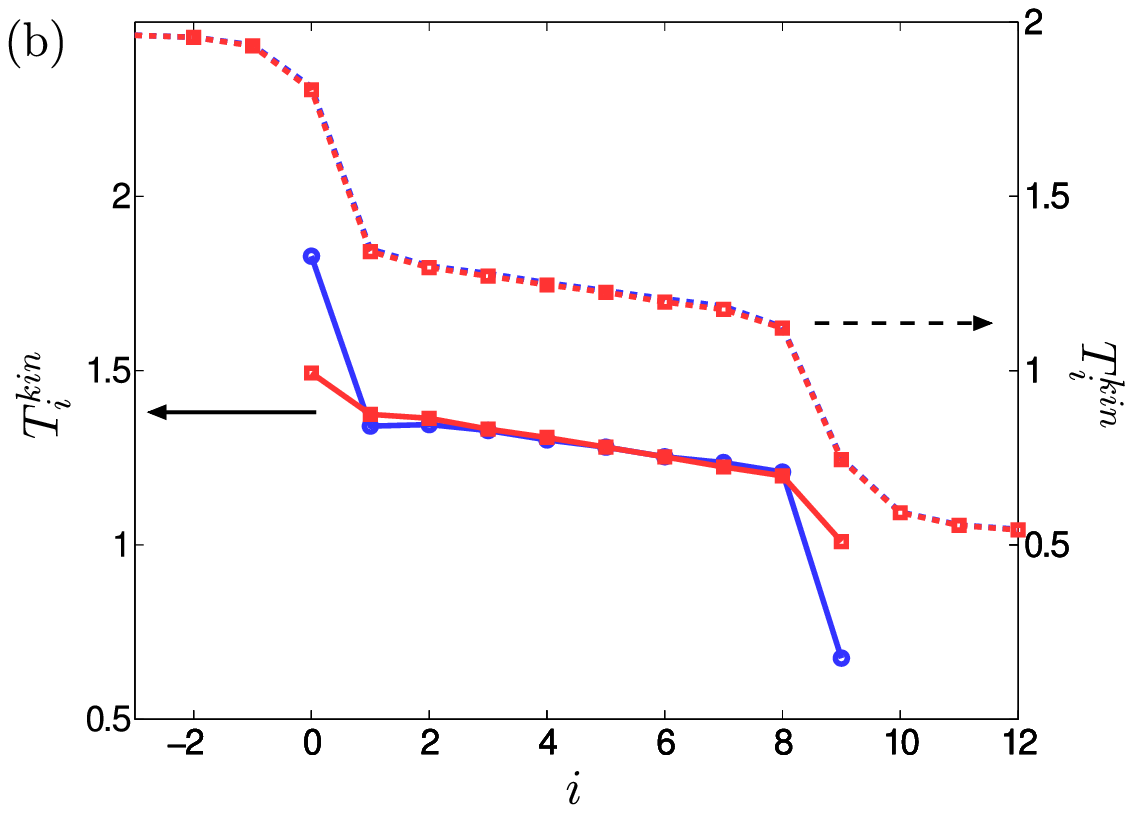}
 \caption{(Color online) Local kinetic temperatures along a chain coupled to the bulk ($L^C=8$, $W^C=1$, $L^R=30$, $W^R=61$, dashed lines) and in a bath-terminated chain of $8$ atoms (solid lines) at (a) low temperature ($T_+=0.20$, $T_-=0.05$) and (b) high temperature ($T_+=2$, $T_-=0.5$), for two Langevin friction parameters $\gamma$. Note that the vertical scale of the temperature profile for the extended structure has an off-set for clarity and the curves for $\gamma=0.5$ and $\gamma=2.0$ lie on top of each other. The error bars $5\delta$ are smaller than the sizes of markers. The insets in (a) are for clarification.}
 \label{fig:tprofchain}
 \end{center}
\end{figure}

One of the motivations of including bulk regions in our simulations is to see how much the bulk and the contact resistance alter the temperature profile and conductivity compared to the classic FPU chain that includes the thermostatted atoms directly at the ends of the chain (Fig. \ref{fig:geom1}(b)). Figure \ref{fig:tprofchain} compares the temperature profiles of chains with $8$ atoms terminated by a thermostatted single atom bath or the extended bulk bath at (a) low temperature and (b) high temperature. To illustrate the dependence of the temperature profiles on the friction parameter $\gamma$, we include the results for $\gamma=0.5$ in addition to $\gamma=2$. In the case of single-atom-bath-termination, $\gamma$ can be considered as a fitting parameter that has no direct relation to the $\gamma$ that we use at the edges of the extended structure (Fig. \ref{fig:geom1}(a)). The atom indexed by $0$ is in a Langevin bath for the single-atom-bath-terminated chain and is the atom closest to the chain in the 
extended structure. 

At low temperature (Fig. \ref{fig:tprofchain}(a)), the temperature gradient in the chain is very small due to nearly ballistic phonon transfer. For the single-atom-bath-terminated chain (solid lines) with $\gamma=2.0$, one can see especially well the local temperature minimum and maximum near the hot and cold ends of the chain, respectively. These features are well-known from the exact solution for the harmonic chain \cite{rieder67} and were suspected to arise from the ''mismatch between the frequencies of the reservoirs and the oscillators''. For the extended structure (dashed lines), on the other hand, the temperature profile is monotonous. Thus the more realistic coupling to the reservoirs seems to eliminate the peculiar nonmonotonous features at the ends of the chain. At higher temperature (Fig. \ref{fig:tprofchain}(b)), scattering increases and a temperature gradient develops inside the chain. In the middle of the chain, the temperature gradients agree quite well for both setups. 

At both high and low temperature, one notices that whereas the temperature profiles with edge-thermalization depend visibly on the Langevin friction parameter $\gamma$, using the extended structure removes this dependence. This suggests that the use of white-noise Langevin dynamics at the edges of the bulk does not produce computational artefacts in the results. Instead, the temperature profiles are determined only by the intrinsic properties of the structure. Curiously, it seems that the boundary jumps in the temperature can be reproduced surprisingly well by choosing $\gamma=2.0$ at the edge of the chain. We have not checked explicitly whether this is accidental or if it could be argued that the value $\gamma=2.0$ describes the spectral properties of square-lattice reservoirs especially well. The value $\gamma=\sqrt{3}/2$ is known \cite{lepri03} to maximize the current flowing through the chain. 

\begin{figure}
 \begin{center}
 \includegraphics[width=8.6cm]{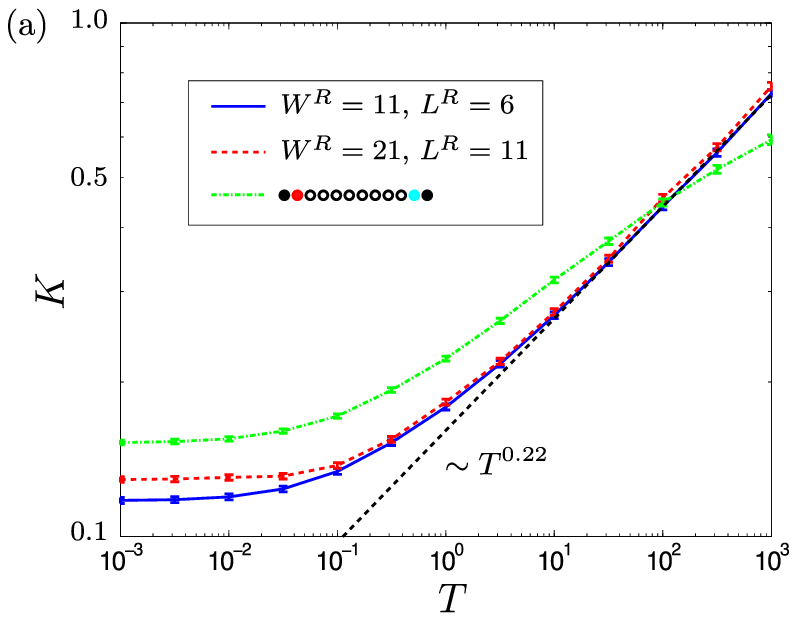}
 \includegraphics[width=8.6cm]{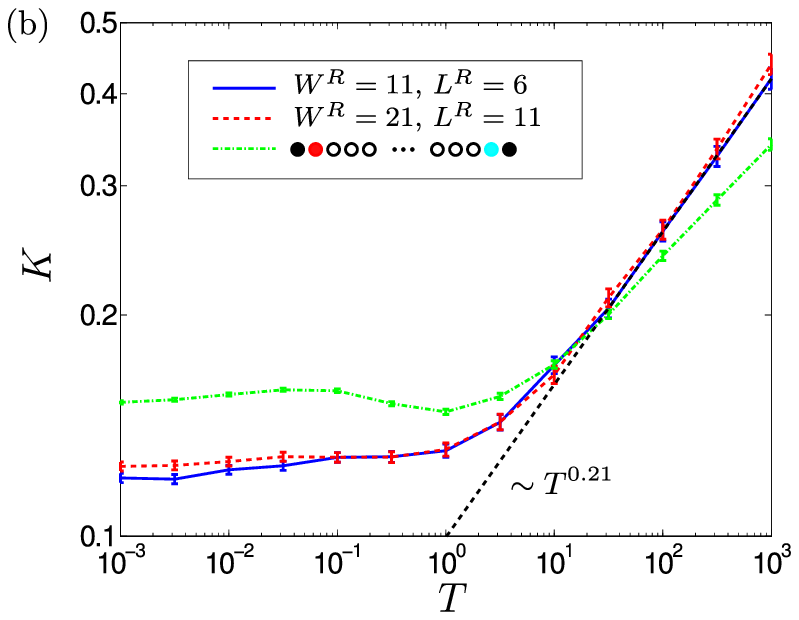}
 \caption{(Color online) Thermal conductance $K$ versus temperature $T$ in chains of length (a) $N=8$ and (b) $N=50$ for two extended structures and the bath-termination. Langevin friction parameter is $\gamma=2$. At high temperature, the thermal conductances of the extended structures scale as $T^{0.22}$ and $T^{0.21}$ for the two chain lengths (black, dashed line). The heat current has been averaged over the length of the chain. The error bars are of length $5\delta$ (Eq. \eqref{eq:varx}).}
 \label{fig:kt1}
 \end{center}
\end{figure}

Figure \ref{fig:kt1}(a) shows the thermal conductance $K(T)$ (Eq. \eqref{eq:KT}) of a short $N=8$ chain connected to two bulk regions of different sizes, compared with the bath-terminated chain of eight atoms. The Langevin friction parameter is set to $\gamma=2$. We set the temperature difference to $\Delta T=2T/9$ as in Ref. \cite{nicolin10}, so that $T_+=10T/9$ and $T_-=8T/9$. At a low temperature, the phonon transport in the system is nearly ballistic and the thermal conductance depends quite strongly on the bulk size, since for a bigger reservoir, there are more vibrational modes. In addition, $K(T)$ is independent of temperature, which follows directly from the classical limit of Landauer--B\"uttiker formula for phonon thermal conductance \cite{angelescu98,rego98}. 

At high temperature, phonon scattering reduces the dependence on the bulk size, which can be seen in Fig. \ref{fig:kt1} as the convergence of the thermal conductance curves at moderately high $T$. Interestingly, the thermal conductance of the bath-terminated chain scales very differently compared to the bulk bath case at high temperatures. A direct fit gives the scaling $K\sim T^{0.22}$ for the extended structure, whereas the bath-thermalized chain has a slower $T$-dependence. This difference is most likely related to the fact that increasing the temperature not only increases the intrinsic thermal conductance of the chain, but also increases the mode populations in the two-dimensional bulk and therefore the total conductance. 

In a longer chain (Fig. \ref{fig:kt1}(b)), the thermal conductance curves look similar to Fig. \ref{fig:kt1}(a) with the exception that conductance momentarily decreases at $T\sim1$ for the bath-terminated structure. This dip has been previously discussed by Nicolin and Segal \cite{nicolin10}, who noted that the dip exists only if coupling to the heat baths is weak and the chain is long enough. The initial decrease of thermal conductance in long chains is most likely related to the $T^{-1}$ dependence of the intrinsic thermal conductivity of FPU chain at low temperature \cite{li07}. The decrease is, however, suppressed by strong coupling to the heat baths.

Similarly to the case of the short chain (Fig. \ref{fig:kt1}(a)), the high-temperature scaling of the conductance is slower for the bath-terminated case. The scaling is now $K\sim T^{0.21}$, where the suggested smaller exponent could be explained by the fact that in a longer chain, the scaling is affected more by the conductance of the chain (slower $T$-dependence) than by the conductance of the bulk (stronger $T$-dependence). 

Let us point out a few important features of Fig. \ref{fig:kt1}. (i) Especially in the bath-terminated case, the magnitude of the conductance depends strongly on the friction parameter $\gamma$, since it determines the ''contact resistance'' between the imaginary thermal bath and the system. In the bulk-terminated case, this bath resistance is supplemented by the interface resistance between the bulk region and the chain, thereby reducing the relative importance of the bath resistance and therefore the dependence of thermal conductance on $\gamma$. (ii) The unbounded increase of the thermal conductance at arbitrarily high energies follows from Ohmic nature of the thermal bath, i.e. the coupling function $\Gamma(\omega)\propto\omega$ for arbitrarily high frequencies \cite{dhar06}. Realistic reservoirs would include a Debye cut-off restricting the occupation of high-energy modes, meaning that scattering would force the conductance to zero at high temperatures. Various thermal baths will be studied in future works.

We have also studied the length-dependence of thermal conductance in FPU chains. There is strong evidence \cite{mai07} that the thermal conductivity $\kappa = JN/\Delta T$ diverges as $N^{1/3}$ in the $N\rightarrow \infty$ limit. We have verified that for very long chains, the bulk termination at the ends of the chain does not affect the observed scaling exponent of $1/3$. This is understandable, since in extremely long systems, the thermal conductance is solely determined by the scattering in the chain and the boundary conditions do not play a role.

\subsection{Discussion}
\label{sec:discussion}
To understand the physics behind our results, some remarks of the simple model based on FPU potentials and square lattice are in order. (i) Our model inherently includes the contact resistance that arises from the mode-dependent coupling between the vibrational modes in the bulk and in the constriction. There is also an artificial contact resistance at the locations of the Langevin bath at the edges of the structure, but making the reservoirs large enough renders this contribution negligible. (ii) Since the equilibrium positions of atoms in our model are fixed, surface reconstruction effects are neglected. Including a realistic interatomic potential is a straightforward extension of our model. (iii) The bulk region contains both phonon-phonon scattering and boundary scattering at the free surface affecting the local phonon density of states near the constriction, meaning that the vibrational properties of the bulk near the constriction are different than deep in the bulk, just as in a real experimental setup. (iv) The two-dimensional scalar FPU model is known to exhibit anomalous transport properties \cite{dhar08}. We do not, however, believe that this affects our qualitative conclusions. Diffusive heat transfer in the bulk regions could be simulated e.g. by breaking momentum conservation using a pinning potential or by extending the model to three dimensions.

To get more extensive or quantitative results, the current study could be extended to several directions. Molecular dynamics simulations allow the calculation of various correlation functions that provide detailed information about spatial variations e.g. in local phonon densities of states or in heat current autocorrelations. The effect of geometry, such as the constriction width and length, also needs to be quantified more closely. Extending the model towards a more realistic model is expected to be relatively straightforward, though computationally demanding.

\section{Conclusions}
\label{sec:conclusions}


In conclusion, we have used the molecular dynamics model with anharmonic FPU potentials to study the phononic heat flow in microscopic constrictions connecting two bulk regions. To study the contact interface effects, we also included large bulk regions in the simulations. We found that (i) in the low temperature region, the temperature profile in the bulk exhibits strong directional patterns following from the mode-dependent coupling between the vibrational modes in the constriction and bulk. At higher temperature, the scattering between vibrational modes increases and smoothens out the directional features. (ii) Directional features and clear interference patterns are also visible in the heat current profile at low temperature. In addition, more heat current was observed to flow at the edges of the constriction than in the center. (iii) Local temperature minima and maxima seen in the bath-terminated atomic chain \cite{rieder67} vanish if the chain is coupled to a larger structure. (iv) The temperature dependence of FPU chain thermal conductance is stronger in the extended structure than in the bath-terminated setup. We expect that our model and results provide new insight into the problem of heat transfer through nanoconstrictions and may help in the design of new thermal materials.

\section{Acknowledgements}
We acknowledge the Finnish IT Center for Science for computer time. The work is in part funded by the MIDE research program of Aalto University.


\bibliographystyle{../physrev}

\end{document}